\documentclass[preprint,superscriptaddress,prd]{revtex4}

\usepackage{amssymb,amsmath,graphicx,amscd,xcolor,dsfont}
\usepackage[colorlinks,citecolor=blue,urlcolor=blue]{hyperref}

\def\E{\mathbf{E}}

\def\P{\mathbf{P}}
\def\x{\mathbf{x}}
\def\k{\mathbf{k}}

\begin{document}
\title{Quantum induced birefringence in nonlinear optical materials}
%
\author{V. A. \surname{De Lorenci}}
\email{delorenci@unifei.edu.br}
\affiliation{Instituto de F\'{\i}sica e Qu\'{\i}mica,    Universidade Federal de Itajub\'a, \\
Itajub\'a, Minas Gerais 37500-903, Brazil}
\author{L. H. \surname{Ford}}
\email{ford@cosmos.phy.tufts.edu}
\affiliation{Institute of Cosmology, Department of Physics and Astronomy, \\
Tufts University, Medford, Massachusetts 02155, USA}

\begin{abstract}
The influence of quantum fluctuations of the electromagnetic field on the propagation of a polarized light wave in a nonlinear dielectric is investigated.
 It is shown that in some cases, the fluctuations couple to the optical nonlinearities of the medium and make its refractive index dependent on the polarization 
 of the propagating wave. As a consequence, two light waves propagating in a same direction but with different polarizations  will have different speeds, 
 so there will be a Kerr-effect birefringence induced by quantum fluctuations. We consider the case of the electromagnetic field in a squeezed vacuum state, where there are
 regions where the mean squared electric field can be negative and the birefringence effect has the opposite sign to the Kerr effect in classical physics.
We  give some  estimates  for the magnitude of these effects, and discuss their possible observability.
\end{abstract}

		
\maketitle

\section{Introduction}
\label{introduction}
An optical medium can be characterized by its polarizability when subjected to an external electromagnetic field. 
In most cases the polarization vector $\P$ can be written in terms of a power series in the electric field $\E$. The coefficients in such expansion are the susceptibilities 
of the material and they encompass the information about the dielectric properties of the medium.  
Particularly, when a probe electromagnetic field  propagates in such media it will experience a refractive index that can be dependent on its polarization state. Hence, 
we can devise a situation where two light signals propagating in the same direction but with different polarizations will have different speeds. This phenomenon of birefringence 
is well known.  It can occur naturally in a material with a fixed optical axis determined by its molecular arrangements, such as a quartz crystal, or artificially in nonlinear media 
where an artificial optical axis is induced by an externally applied electric field.  In the latter case, the induced optical axis only survives in the presence of the external field. 
The first observations of artificial electric and magnetic birefringence were reported more than a century ago~\cite{kerr1875,cotton1905}. 
Both natural and induced effects can occur in a same system, leading to nontrivial mechanisms~\cite{teodoro2004} for  controlling birefringence.

Birefringence is also possible in nonlinear theories of the electromagnetism, where the corresponding Lagrangian is a nonlinear function of the electromagnetism invariants. 
The most famous case being  quantum electrodynamics (QED). Here vacuum birefringence due to vacuum polarization was predicted to 
occur~\cite{heisenberg1936,weisskopf1936} in a regime of strong fields even before the complete formulation of the theory.
The  light propagation in various nonlinear theories, including birefringence, has been investigated by many authors (see, for example, 
Refs.~\cite{birula1970,adler1971,dittrich1998,delorenci2000,obukhov2002}).
Recently, evidence for vacuum birefringence was reported in optical-polarimetry measurement of an isolated neutron star \cite{mignani2017}, which may arise
from QED effects in a strong magnetic field. Birefringence is also predicted by some models of quantum gravity~\cite{GP99}.

In this paper we investigate another possible way of producing birefringence in a nonlinear optical medium. Classically, an artificial optical axis induced by 
an external electric field disappears as soon as the field is turned off. However, quantum mechanics teaches us that there is no way of completely turning off 
a fundamental field. There will always remain its quantum fluctuations. Our approach consists of preparing a background electric field in a squeezed vacuum state, 
and studying its effect on probe electromagnetic waves propagating in the medium whose nonlinearities are activated by the background field.

In Sec.~\ref{birefringence},   we discuss how
the expectation value of the squared background electric field couples to the third order polarizability of the medium and can produce 
birefringence phenomena. We exhibit a specific physical system where this effect is expected to occur, and give estimates for the magnitude of the 
difference in the propagation speeds of different polarization states.
Next, in Sec. \ref{effects} we study the implications of the sign of the expectation value of the squared electric field on the wavelength  modulation of a probe field. 
In Sec. \ref{comparing} we compare the expectation value of the squared electric field in a squeezed vacuum state with that in a coherent state. Particularly, it is shown that, 
in the limit of small numbers of photons per mode, the peak magnitude of this quantity can be larger in the squeezed  state than in the coherent state. 
The effects of thermal fluctuations are studied in Sec.~\ref{Casimir}, as are the Casimir vacuum fluctuations near a reflecting boundary.
Final remarks are given in Sec. \ref{final}.

We use Lorentz-Heaiviside units with $\hbar=c=1$, unless otherwise  stated. Moreover, we set $\varepsilon_0 \approx 8.85 \times 10^{-12}\, {\rm C^2/(N \,m^2)} =1$,  which leads to 
$1 V \approx 1.67\times 10^7 {\rm m}^{-1}$.

\section{Birefringence induced by quantum fluctuations}
\label{birefringence}
We start by studying the propagation of an electric field perturbation, $\E^{\tt p}$, the probe field, in a nonlinear optical medium whose nonlinearities are 
activated by an externally applied background electric field, $\E^0$. Let the polarization of the background field be in the $x$-direction, so that its $i$-th component is 
$E_i^0 = E^0(t,\x)\delta_{ix}$. Moreover, suppose the probe field propagates in the $z$-direction, and that the conditions 
$\| E_i^0\| \gg \| E_i^{\tt p}\|$ and $\| \nabla E_i^0\| \ll \| \nabla E_i^{\tt p}\|$ are satisfied. In this case, Maxwell's equations lead~\cite{ford2013}  to the wave equation 
\begin{equation}
\frac{\partial^2 E_i^{\tt p}}{\partial z^2} - \sum_{j=1}^3\biggl[ \delta_{ij} + \chi_{ij}^{(1)} + 2\chi_{i(jx)}^{(2)} E^0 
+ 3\chi_{i\{jxx\}}^{(3)} (E^0)^2 + \cdots \biggl] \frac{\partial^2 E_j^{\tt p}}{\partial t^2} = 0    \,,
\label{wave-eq}
\end{equation}
where $\chi_{ij}^{(1)}$ denote the components of the linear susceptibility tensor, while the coefficients $\chi_{ijk}^{(2)}$ and $\chi_{ijkl}^{(3)}$ denote 
the components of the second- and third-order nonlinear susceptibility tensors, respectively~\cite{boyd2008} . We define 
$\chi_{i(jk)}^{(2)} \doteq (1/2)\left(\chi_{ijk}^{(2)}+\chi_{ikj}^{(2)}\right)$ and $\chi_{i\{jkl\}}^{(3)} \doteq (1/3)\left(\chi_{ijkl}^{(3)}+\chi_{iklj}^{(3)}+\chi_{iljk}^{(3)}\right)$. 
Notice that the term between square-brackets in the wave equation is the square of the refractive index measured by the probe field $E_i^{\tt p}$. 

In the limit that $E^0  = 0$, any birefringence will arise from the linear susceptibility tensor, $\chi_{ij}^{(1)}$. Here we assume that $\chi_{xx}^{(1)}=\chi_{yy}^{(1)}$, so no natural
birefringence occurs for propagation in the $z$-direction. Let the probe field be polarized along the $x$-direction, so $E^{\tt p_1}_i = E^{\tt p_1}(t,z) \delta_{ix}$. From
Eq. (\ref{wave-eq}) we see that the probe field experiences an effective squared refractive index of
\begin{equation}
n_{\tt p1}{}^2 =  n_0{}^2+ 2\chi_{xxx}^{(2)} E^0 + 3\chi_{xxxx}^{(3)} \left(E^0\right)^2,
\label{np1}
\end{equation}
where we defined the ordinary refractive index $n_0{}^2 \doteq 1 + \chi_{xx}^{(1)} = 1 + \chi_{yy}^{(1)}$. On the other hand,  if the probe field is prepared with polarization 
in the $y$-direction so, $E^{\tt p_2}_i = E^{\tt p_2}(t,z) \delta_{iy}$, it will experience a squared refractive index of
\begin{equation}
n_{\tt p2}{}^2 =  n_0{}^2 + 2\chi_{y(yx)}^{(2)} E^0 + 3\chi_{y\{yxx\}}^{(3)}\left(E^0\right)^2,
\label{np2}
\end{equation}
which is generally distinct from $n_{\tt p1}{}^2$, depending on the nonlinear electric properties of the optical material. 
The magnitude of the induced birefringence effect can be measured by the fractional difference $\delta n \doteq (n_{\tt p1} - n_{\tt p2})/{n_0}$. Note that the power expansion 
in Eq.~(\ref{wave-eq}) assumes that $\big|\chi_{ijkl}^{(3)} (E^0)^2\big| \ll \big|\chi_{ijk}^{(2)} E^0\big| \ll \big|\chi_{ij}^{(1)} \big|\approx 1$. Additionally, we assume that 
$\left(\chi_{ijk}^{(2)}/n_0{}^2\right)^2 \ll \big|\chi_{ijkl}^{(3)}/n_0{}^2\big|$.
With these approximations we obtain  
\begin{eqnarray}
\delta n \approx \left(\frac{\chi_{xxx}^{(2)}-\chi_{y(yx)}^{(2)}}{n_0{}^2}\right)E^0 +\frac{3}{2}\left(\frac{\chi_{xxxx}^{(3)}-\chi_{y\{yxx\}}^{(3)}}{n_0{}^2}\right)\left(E^0\right)^2.
\label{delta}
\end{eqnarray}

The first term on the right hand side of the above equation gives an effect linear in $E^0$  known as Pockels effect. This effect does not occur in centrosymmetric systems, 
for which the second order polarizability must vanish due to spatial inversion symmetry. The second term in Eq. (\ref{delta}) gives the Kerr effect, which is  birefringence 
quadratic in the applied background field $E^0$. Now we consider the situation where the background field undergoes quantum fluctuations around a zero mean value.
That is, we now describe the electric field as a quantum operator ${\hat E}^0 (t,{ \x})$, whose  expectation value is zero, $\langle {\hat E}^0 (t,{ \x}) \rangle = 0$. 
The fractional difference is now a quantum observable $\delta {\hat n}$, whose  expectation value is given by 
\begin{equation}
\langle \delta {\hat n} \rangle = \frac{1}{2n_0{}^2}\left(3\chi_{xxxx}^{(3)} - \chi_{yyxx}^{(3)}- \chi_{yxxy}^{(3)}- \chi_{yxyx}^{(3)}\right)\langle :\!({{\hat E}^0})^2 \!: \rangle\,.
\label{quantum-delta}
\end{equation}
In this expression we normal ordered the squared electric field, $ :\!({{\hat E}^0})^2 \!: \;= ({{\hat E}^0})^2 - \langle 0|({{\hat E}^0})^2 |0 \rangle $, because the effect 
we are investigating is related to the change in the expectation value of the field as compared to the empty space vacuum state $|0\rangle$. 
Compared to the classical expression [Eq. (\ref{delta})] we see that, when $ \langle \hat E^0 \rangle  = 0$  , there will be no quantum induced birefringence effect 
associated with second order polarizability. The first contribution to the effect appears in the third order term, which is coupled to the square of the background electric field. 
This is a type of Kerr effect induced by quantum fluctuations of the electric field. Note that this effect is very different from the birefringence due to vacuum polarization
mentioned in the previous section.

A possible scenario where birefringence induced by quantum fields appears is when the background electric field is prepared in a multimode squeezed vacuum state. 
Such a state can be described by applying the squeeze operators for each wave mode $\k$, 
$\hat S(\zeta_\k) = \exp\left\{\frac{1}{2}\left[\zeta^*_\k (\hat a_\k)^2 - \zeta_\k (\hat a^\dagger_\k)^2\right] \right\}$, on the  vacuum state 
$|0\rangle$ as $|\psi\rangle = \Pi_\k \hat S(\zeta_\k) |0\rangle \doteq  \Pi_\k |\zeta_\k\rangle$. Here $\hat a_\k$ and $\hat a_\k^\dagger$ are the 
annihilation and creation operators, respectively, and $\zeta_\k$ denotes the complex squeeze parameter whose amplitude and phase are given 
by $q_\k$ and $\eta_\k$, respectively. 
If we expand the electric field operator in normal modes, use the commutation relations for the creation and annihilation operators, and take the expectation value 
of the square of the electric field in a multimode squeezed state $|\psi\rangle$ we obtain that \cite{ford2013}
%
$\langle :\!({{\hat E}^0})^2 \!: \rangle_{{}_\psi} =  \frac{1}{V}\,  \sum_{\k} \omega \,
\sinh q_\k\, \left[ \sinh q_\k +  \cosh q_\k\, \cos(2\omega t -2\k \cdot \x-\eta_\k) \right]$,
%
where $V$ indicates the quantization volume, and $\omega = \omega(\k)$ is the frequency related to mode $\k$.

Passing to the continuum limit, which corresponds to a large quantization volume, and specializing to the case where the excited modes 
of the field are peaked about the angular frequency $\omega = \Omega$ and propagate in the $y$-direction, $\k = (0,k,0)$, with a small but nonzero fractional 
bandwidth $\Delta k/k \ll 1$, we find
\begin{equation}
\langle :\!({{\hat E}^0})^2 \!: \rangle_{{}_\psi} = \frac{1}{4 \pi^2}\,   \Omega \, k^2\,\Delta k \, \Delta \theta \,  \sinh q  \left[ \sinh q+  \cosh q \cos(2\Omega t -2ky-\eta)\right]\,,
\label{E2b}
\end{equation}
where $\Delta\theta$ denotes the angular spread of the field modes around the $y$-direction, and $q_\k \approx q$ is constant for the excited modes.
For a collimated beam, we would require that $\Delta\theta \ll 1$. 
However, the expectation value of the squared electric field in a wave guide can take a form similar to Eq.~(\ref{E2b}), but with $\Delta\theta \approx 1$. 
(See Eq.~(42) in Ref.~\cite{DF19}.) Notice that, depending of the values 
that the parameter $q$ takes, negative values of $\langle :\!({{\hat E}^0})^2 \!: \rangle_{{}_\psi}$ are possible. Such behavior is associated with subvacuum fluctuations 
and is a purely quantum result not found in classical physics. Because $\langle  0|:\!({{\hat E}^0})^2 \!: |0 \rangle = 0$, negative values on the mean squared electric field
arise when there is suppression of the quantum fluctuations below the vacuum level. The squeezed states which create the maximal subvacuum effect were constructed
in Ref.~\cite{KF18}, and some of the observable consequences of the effect were treated in Ref.~\cite{DF19}.  The regions of negative expectation value of 
$\langle :\!({{\hat E}^0})^2 \!: \rangle_{{}_\psi}$ lead to an increase in the speed of the probe pulse, which is analogous to the effects of negative energy density
in general relativity~\cite{bessa2014}.

\begin{figure}
\includegraphics[scale=0.6]{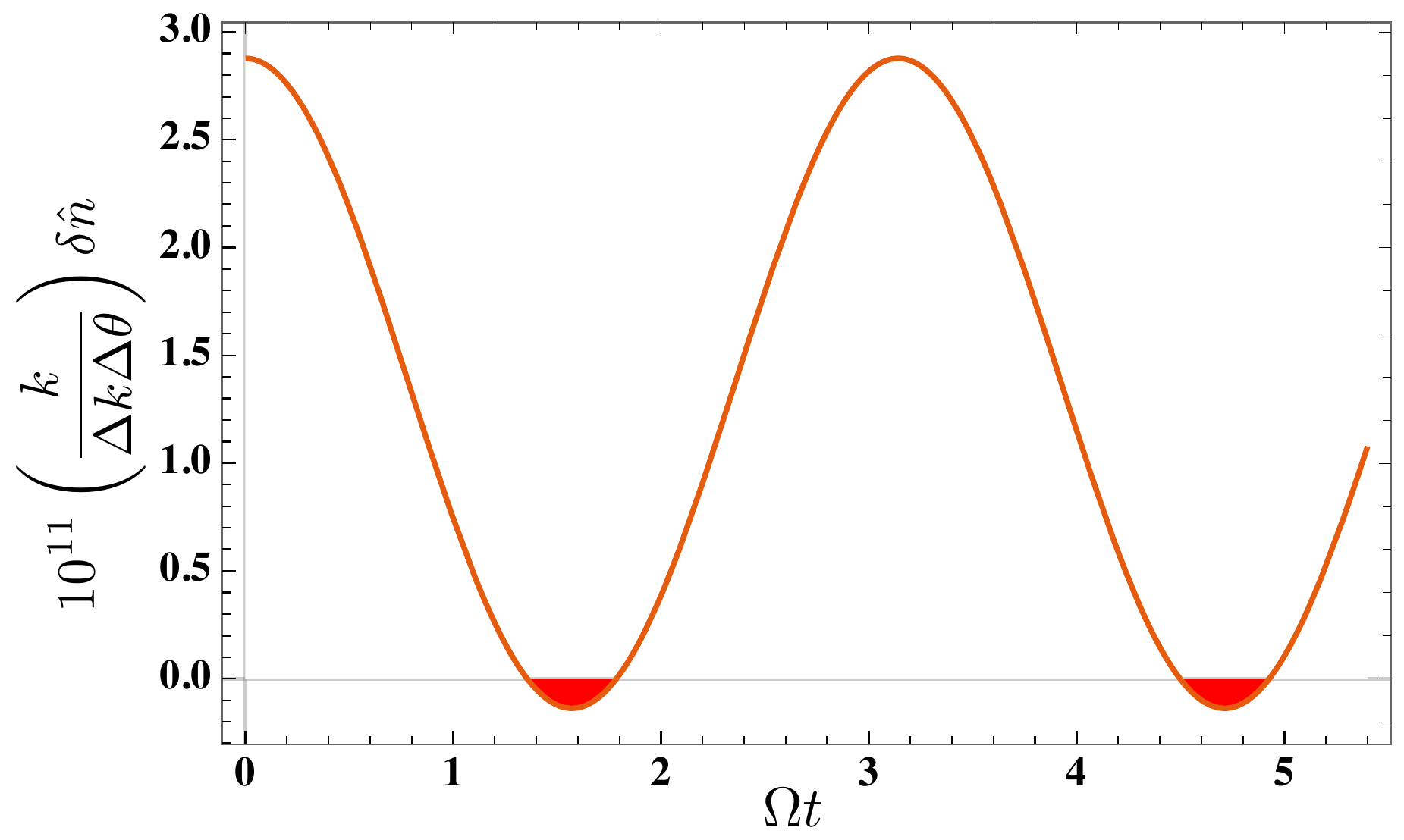}
\caption{Behavior of $\delta {\hat n}$ described by Eq. (\ref{Deltan}) with the choice of phase $\eta = -2ky_0$. Notice that the subvacuum phenomenon occurs periodically 
but within a small interval of time.}
\label{fig1}
\end{figure}

An example of a material that could exhibit quantum induced birefringence is the chalcopyrite crystal $\mbox{CdGeAs}_2$ (cadmium germanium arsenide). 
This is a semiconductor crystal 
with low dispersion  for wavelengths in the range $8 \mu{\rm m} \lesssim \lambda \lesssim 12 \mu{\rm m}$, where the ordinary refractive index is 
$n_0 \approx 3.5$ \cite{boyd1972}. The relevant third order susceptibility coefficients for this material at $\lambda = 10.6\mu{\rm m} $ are ~\cite{charra2000} 
$\chi_{xxxx}^{(3)} \approx 72800\times 10^{-22}{\rm m}^2{\rm V}^{-2}$ and $\chi_{xxyy}^{(3)}\approx -14000\times 10^{-22}{\rm m}^2{\rm V}^{-2}$. With these results we obtain 
$\langle \delta {\hat n} \rangle_{{}_\psi} \approx 3.39 \times 10^{-9} \langle :\!({{\hat E}^0})^2 \!: \rangle_{{}_\psi} (\mu{\rm m})^{4} $. Now, in order to estimate 
$\langle :\!({{\hat E}^0})^2 \!: \rangle_{{}_\psi}$ we use the same assumptions used in Ref. \cite{bessa2014}. First, let us assume a squeezed state presenting a squeezing 
level of $10 {\rm db}$~\cite{vahlbruch2008},  which corresponds to  $q=1.5$. Furthermore, setting the wavelength of the background field to be 
$\lambda = 2\pi/k = 10.6\mu{\rm m}$, we obtain $(\Omega k^3)/(4\pi^2) \approx 8.93\times 10^{-4} (\mu{\rm m})^{-4}$. Introducing these results in Eq.~(\ref{E2b}) yields
\begin{equation}
\langle :\!({{\hat E}^0})^2 \!: \rangle_{{}_\psi} \approx 4.05\times 10^{-3} \left( \frac{\Delta k \, \Delta \theta}{k}\right)\big[ 1+ 1.1 \cos(2\Omega t -2ky-\eta)\big] (\mu{\rm m})^{-4} \,.
\label{E2c}
\end{equation}
With the above results, we find that the magnitude of the birefringence effect induced by the expectation value of the squared electric field in the squeezed vacuum state is 
\begin{equation}
\langle \delta {\hat n} \rangle_{{}_\psi} \approx  1.37 \times 10^{-11} \left( \frac{\Delta k \, \Delta \theta}{k}\right)\big[ 1+ 1.1 \cos(2\Omega t -2ky-\eta)\big]\,.
\label{Deltan}
\end{equation}
If we fix a spatial position $y=y_0$ and measure $\delta {\hat n}$, we will find that this quantity will alternate in time between positive and negative values, as depicted in 
Fig.~\ref{fig1}.  As it is clear, subvacuum effects may occur whenever the term between square brackets has a negative value, i.e., whenever $1.1 \cos(2\Omega t -2ky-\eta) <-1$, 
as shown in the filled regions in the figure. 

As we have seen, the system will be affected by positive and negative effects of the  fluctuations. The maximum magnitude of the positive effect of fluctuations is 
$\langle \delta {\hat n} \rangle_{{}_\psi} \approx  2.88\times 10^{-11} (\Delta k/k) \Delta \theta$, while for the negative effect we have 
$\langle \delta {\hat n} \rangle_{{}_\psi} \approx  -1.37\times 10^{-12} (\Delta k/k)\Delta \theta$. If we accept that $(\Delta k/k) \Delta \theta$ can be arranged to be as 
large as $10^{-4}$, we could have a potentially measurable effect.

Closing this section, we should notice that many of techniques of detection in optics are based on time averaging procedures. In such cases the last term in 
Eq. (\ref{Deltan}) would not contribute to the measured effect, which means the existence of subvacuum fluctuations would not be noticed. 
However, in Ref.~\cite{DF19} a proposal for amplifying the subvacuum effect was given. It involves a
localized probe wavepacket which fits in the region where $\langle :\!({{\hat E}^0})^2 \!: \rangle < 0$ and propagates with this region. In addition, the existence
of regions where $\langle :\!({{\hat E}^0})^2 \!: \rangle <0$ can lead to subtle changes in the probe field, as was discussed in Ref.~\cite{DF19} and in 
the next section.

\section{Effects of the sign of the squared electric field upon the frequency modulation of a probe field}
\label{effects}
Let us now discuss further consequences of the sign of $\langle :\!({{\hat E}^0})^2 \!: \rangle$ on the behavior of the probe field. Return to the wave equation,
 Eq.~(\ref{wave-eq}), consider the case of a probe field where $E_i^{\tt p} = E^{\tt p}(t,z)\delta_{xi}$, and take the expectation value in a state where 
 $\langle {\hat E}^0 (t,{ \x}) \rangle = 0$.  The resulting equation describes propagation of the probe field at an effective velocity of
\begin{equation}
v_{\tt eff} = \frac{v_0}{ 1+f(t,\x)}\;,
\label{veff}
\end{equation}
where  $v_0 \doteq 1/n_0$ is the phase velocity of the wave in the absence of the background field, and we have defined 
$f(t,\x) = 3 v_0{}^2 \chi_{xxxx}^{(3)} \langle :E^0(t,\x)^2: \rangle$. 
As we have seen, $\langle :\!({{\hat E}^0})^2 \!: \rangle$ can be positive or negative.

\begin{figure}
\includegraphics[scale=0.75]{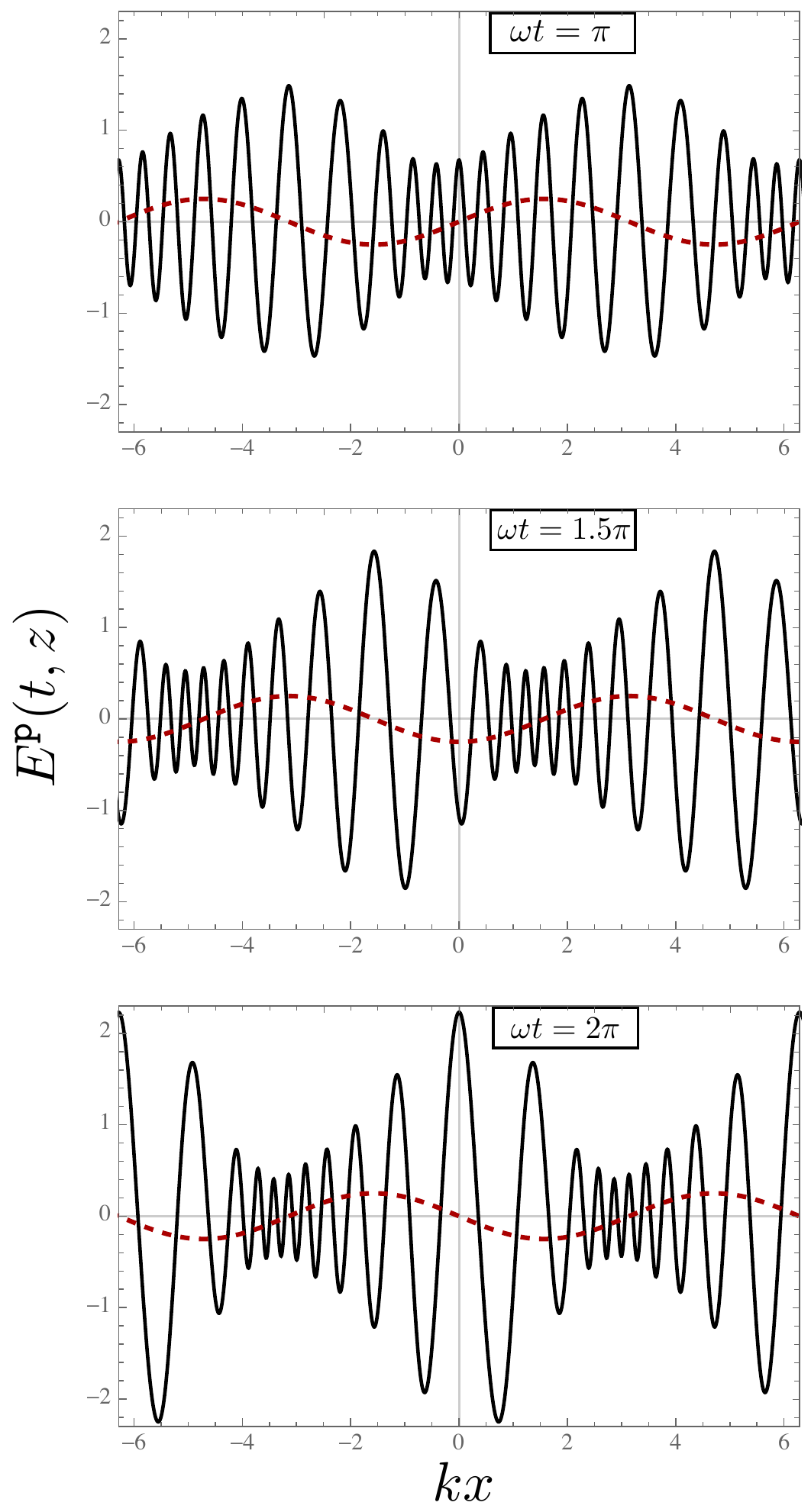}
\caption{Behavior of the probe field $E^{\tt p}$  as function of distance $z$ for different instants of time. The dashed curve describes the function 
$ f(t,\x)$, which gives the contribution to the phase velocity of the probe field due to the quantum fluctuations.}
\label{fig2}
\end{figure}
Now suppose the probe field is described by a plane wave such that $E_j^{\tt p} \sim {\rm e}^{i(kz-\omega t)}\delta_{xj}$. In this case $v_{\tt eff} = \omega / k$ and, 
if we treat $\omega$ as a constant, it follows that $k = k(t,z) = 2\pi/\lambda(t,z)$. Hence, in the region where $\langle :\!({{\hat E}^0})^2 \!: \rangle <  0$ we will have 
$v_{\tt eff} > v_0$, which means that the wavenumber $\lambda$ increases. In fact, we can view the changes in $v_{\tt eff}$ as due to the changes in $\lambda$ with 
fixed $\omega$. Thus, a decrease (increase) in $\langle :\!({{\hat E}^0})^2 \!: \rangle$ causes $\lambda$ to increase (decrease) and hence $v_{\tt eff}$ to increase (decrease).
This effect is numerically studied in Fig.~\ref{fig2}, where we assumed a simple model for which $f(t,\x)  = -0.25 \sin (kz - \omega t)$ and numerically integrated the
wave equation, Eq.~(\ref{wave-eq}). 

The probe field solution of the wave equation was obtained by using the initial conditions  $E^{\tt p}(t=0,z) = \cos(10 z)$ and 
$(dE^{\tt p}/dt)(t=0,z) = 10\sin(10 z)$.  
As  can be seen in Fig.~\ref{fig2}, as $t$ increases, both $E^{\tt p}(t,z)$ and $\langle :\! \hat f(t,\x) \!: \rangle$ move to right. The region of most rapid oscillation 
of $E^{\tt p}(t,z)$ seems to be somewhat ahead of the region where $f(t,\x) $ is minimum. Moreover, the amplitude of $E^{\tt p}(t,z)$ 
nicreases where its wavelength increases. This is another way to visualize the results in Sec.~III of Ref.~\cite{DF19}, where it was shown that the presence
of $\langle :\!({{\hat E}^0})^2 \!: \rangle < 0$  regions leads to side bands in the frequency spectrum of the probe pulse.

\section{Comparing the effects of squeezed and coherent states in the limit of small occupation number}
\label{comparing}
Consider a plane wave mode propagating in the $y$-direction, with wave vector $\k = (0,k,0)$. The contribution of this mode to the background electric field operator
is $\hat{E}^{0} = E^0\left(\hat a_\k {\rm e}^{i\varphi} +\hat a_\k^\dagger {\rm e}^{-i\varphi} \right)$, with $\varphi \doteq ky -\Omega t$. 
Suppose that this mode is in a coherent state, described by the state vector $|{\cal Z} \rangle$ such that $ \hat a_\k | {\cal Z}\rangle  = {\cal Z} | {\cal Z}\rangle$, where 
${\cal Z}$ is assumed to be real. The expectation value of the normal ordered squared electric field operator 
in this state  is 
\begin{equation}
\langle :\!\!({{\hat E}^0})^2 \!\!: \rangle_{{}_{\cal Z}} \doteq  \langle {\cal Z}| :\!\!({{\hat E}^0})^2 \!\!: |{\cal Z}\rangle  = (2{\cal Z}E^0\cos \varphi)^2 = (\langle \hat E^0 \rangle_{{}_{\cal Z}})^2.
\label{coh}
\end{equation}
Furthermore, as the mean number of photons is given by $\langle \hat n \rangle_{{}_{\cal Z}} = \langle \hat a_\k^\dagger \hat a_\k \rangle_{{}_{\cal Z}} = {\cal Z}^2$, it follows that 
$ \langle :\!\!({{\hat E}^0})^2 \!\!: \rangle_{{}_{\cal Z}}  = (2E^0\cos \varphi)^2 \langle \hat n \rangle_{{}_{\cal Z}}$. Now, taking the time average of these expectation values, 
we obtain  $\overline{\langle \hat E^0 \rangle}_{{}_{\cal Z}} = 0$, and
\begin{equation}
\overline{\langle :\!\!({{\hat E}^0})^2 \!\!: \rangle}_{{}_{\cal Z}} = 2(E^0)^2 \langle \hat n \rangle_{{}_{\cal Z}} \propto \langle \hat n \rangle_{{}_{\cal Z}}.
\end{equation}

If we repeat the above procedure but  prepare this mode in the squeezed vacuum state $|\zeta \rangle = \hat S(\zeta)|0\rangle$,  we obtain
\begin{equation}
\langle :\!\!({{\hat E}^0})^2 \!\!: \rangle_{\zeta} \doteq  \langle \zeta| :\!\!({{\hat E}^0})^2 \!\!: |\zeta\rangle  = (2E^0)^2 \sinh q  \left[ \sinh q +  \cosh q \cos(2\varphi)\right],
\end{equation}
where we have set the phase of the squeeze parameter so that $\zeta = q {\rm e}^{i\eta} = q$.
Time-averaging this result and using that $\langle \hat n \rangle_{\zeta} = \langle \hat a_\k^\dagger \hat a_\k \rangle_{\zeta} = \sinh^2 q$, yields
 \begin{equation}
\overline{\langle :\!\!({{\hat E}^0})^2 \!\!: \rangle}_{\zeta} = 2(E^0)^2 \langle \hat n \rangle_{\zeta} \propto \langle \hat n \rangle_{\zeta}.
\end{equation}
In both cases,  we find that the time averaged expectation value of the squared electric field 
is proportional to the mean number of photons in the corresponding state. 

Now we wish to investigate the regime of small occupation number $\langle n \rangle \ll 1$. Equation~(\ref{coh}) shows that 
$ \langle :\!\!({{\hat E}^0})^2 \!\!: \rangle_{{}_{\cal Z}} \propto \langle \hat n \rangle_{{}_{\cal Z}}$ still holds in this limit. 
However, in the case of a squeezed vacuum state, small occupation number implies $\cosh q \approx 1$ and $q \approx \sinh q = \sqrt{\langle \hat n \rangle_{\zeta}}$, so that
$\langle :\!\!({{\hat E}^0})^2 \!\!: \rangle_{\zeta}  \propto \sqrt{\langle \hat n \rangle_{\zeta}}\cos(2\varphi)$. 
Thus, in the limit of small occupation number, it takes a smaller mean number of photons to produce a given peak of $\langle :\!\!({{\hat E}^0})^2 \!\!: \rangle$ 
in a squeezed vacuum state than in a coherent state. Note that in this limit $\langle :\!\!({{\hat E}^0})^2 \!\!: \rangle_{\zeta}$ oscillates symmetrically around zero in time,
while  $ \langle :\!\!({{\hat E}^0})^2 \!\!: \rangle_{{}_{\cal Z}} \geq 0$.

\section{Thermal and Casimir Effects}
\label{Casimir}

We have been considering birefringence effects in fluctuating electric fields, where the ensemble average value of the electric field vanishes, but that of its
square does not. This situation can also arise in a thermal state or in the Casimir effect. Recall that in Lorentz-Heaviside units, the mean energy density
in the electromagnetic field is
 \begin{equation}
 \langle U \rangle = \frac{1}{2} \left(\langle E^2 \rangle  +\langle B^2 \rangle \right) \,,
 \label{eq:U}
 \end{equation}
and in a thermal state we expect the mean squared electric and magnetic fields to be equal, so $\langle B^2 \rangle = \langle E^2 \rangle$, and 
$\langle E^2 \rangle = \langle U \rangle$. This allows us to use well known results for the energy density in black body radiation to write
 \begin{equation}
 \langle E^2 \rangle = 1 \, {\rm \mu m}^{-4}\, \left( \frac{T}{2600 K} \right)   \,,
\label{eq:thermal}
 \end{equation}
where $T$ is the temperature. We can see from this expression that at any temperature below the melting point of most materials, the mean squared electric
field is small compared to that easily attained in squeezed vacuum states, so thermal effects can generally be neglected in the models which we treat.

However, in the Casimir effect the mean squared electric  field can be large enough to produce a potentially observable birefringence effect. We may estimate
this effect by considering a perfectly reflecting plate in vacuum, where at a distance $z$ from the plate, we have
\begin{equation}
\langle E^2 \rangle  = \frac{3}{16 \pi^2\; z^4}\,.
 \label{eq:perfect}
 \end{equation}
Although the plate defines a preferred spatial direction, $\langle E^2 \rangle$ is isotropic in the case, so 
$\langle E_x^2 \rangle = \langle E_y^2 \rangle = \langle E_z^2 \rangle =  \langle E^2 \rangle/3$. This result may be derived from the renormalized
field strength two point function given in Ref~\cite{BM69}.
Equation~(\ref{eq:perfect}) is a good approximation so long as $z \agt \lambda_P$, the plasma wavelength of the material in the plate. 
Although $\langle E^2 \rangle>0$
in this case, the mean squared magnetic field is negative, $\langle B^2 \rangle  = -\langle E^2 \rangle<0$. If $z \alt \lambda_P$, then~\cite{SF02}
\begin{equation}
\langle E^2 \rangle  \approx \frac{\sqrt{2}}{16 \, \lambda_P\, z^3}\,.
 \label{eq:plasma}
 \end{equation}
Thus if $z \alt 1 {\rm \mu m}$, then values of $\langle E^2 \rangle \agt 1 \, {\rm \mu m}^{-4}$ may be attainable, which are comparable to the values which can be
found in squeezed vacuum states.

Here we have used the idealized case of vacuum outside of a reflecting boundary for an order of magnitude estimate. If a nonlinear material is placed near a reflecting
plate, we can expect the effects to be of the same order as in vacuum, but a more detailed calculation is needed for a better estimate. In all of the case we have
discussed $\langle E^2 \rangle>0$, so the sign of the birefringence effect will be the same as in a classical field. It is of interest to search for Casimir geometries
which can produce regions where $\langle E^2 \rangle<0$, and hence demonstrate a time independent subvacuum effect.

\section{Summary and Conclusions}
\label{final}

We have investigated birefringence induced by fluctuating electric fields. In quantum states in which the mean field vanishes, but the mean square does not, then an induced Kerr effect can occur, where different polarization states propagate at different speeds. We have paid particular attention to the case of squeezed vacuum states, where a subvacuum effect can occur in which the expectation value of the squared electric field becomes negative, below the vacuum value. In this case, the birefringence effect has the opposite sign from that found in classical electric fields.  We have given some estimates of the likely magnitude of this effect. Although it is small, its eventual observation may be possible. 

We have also estimated the effects due to thermal fluctuations and quantum fluctuations in a Casimir vacuum state near a boundary. The thermal effect seems to be very small at room temperature compared to the effects of quantum fluctuations of the electric field. In simple geometries, the Casimir effect leads to a positive expectation value of the squared electric field. It is not yet known whether a negative expectation value can be produced by Casimir vacuum fluctuations.

Here we have dealt with a model using nonlinear dielectric materials. However, as was discussed in Ref.~\cite{bessa2014}, these models can be a reasonable description for effects which are expected to arise in semiclassical and quantum gravity. Thus the study of birefringence in quantum gravity models, which was begun in Ref.~\cite{GP99}, may be worthy of further investigation.

\begin{acknowledgments}
This work was supported in part by the Brazilian agency {\em Conselho Nacional de Desenvolvimento Cient\'{\i}fico e Tecnol\'ogico}  (grant CNPq-302248/2015-3), and by the National Science Foundation (grant PHY-1607118).
\end{acknowledgments}


\begin{thebibliography}{99}

\bibitem{kerr1875}
J. Kerr, 
A new relation between electricity and light: dielectrified media birefringent,
\href{https://doi.org/10.1080/14786447508641302}{Philos. Mag. {\bf 50}, 337 (1875)}.

\bibitem{cotton1905}
A. Cotton and M. Mouton, 
{\em Sur la bir\'efringence magn\'etique,} 
Compt. Rendu. {\bf 141}, 349 (1905).

\bibitem{teodoro2004}
V. A. De Lorenci, R. Klippert, and D. H. Teodoro,
Birefringence in nonlinear anisotropic dielectric media,
\href{https://doi.org/10.1103/PhysRevD.70.124035}{Phys.\ Rev. {\bf D 70}, 124035 (2004)}
[\href{https://arxiv.org/abs/gr-qc/0603042}{arXiv:gr-qc/0603042}].

\bibitem{heisenberg1936}
W. Heisenberg and H. Euler, 
{\em Folgerungen aus der Diracschen theorie des positrons},
\href{https://doi.org/10.1007/BF01343663}{Z. Phys. {\bf 98}, 714 (1936)}.

\bibitem{weisskopf1936}
V. Weisskopf,
 {\em \"Uber die elektrodynamik des vakuums auf grund der quantentheorie des elektrons},
Mat.-Fys. Medd. K. Dan. Vidensk. Selsk. {\bf 14}, n. 6 (1936).

\bibitem{birula1970}
Z. Bialynicka-Birula and I. Bialynicki-Birula, 
Nonlinear effects in quantum electrodynamics: photon propagation and photon splitting in an external field,
\href{https://doi.org/10.1103/PhysRevD.2.2341}{Phys. Rev. D {\bf 2}, 2341 (1970)}.

\bibitem{adler1971}
S. L. Adler, 
Photon splitting and photon dispersion in a strong magnetic field,
\href{https://doi.org/10.1016/0003-4916(71)90154-0}{Ann. Phys. {\bf 67}, 599 (1971)}.

\bibitem{dittrich1998}
W. Dittrich, and H. Gies, 
Light propagation in nontrivial QED vacua,
\href{https://doi.org/10.1103/PhysRevD.58.025004}{Phys.\ Rev. {\bf D 58}, 025004 (1998)}.

\bibitem{delorenci2000}
V. A. De Lorenci, R. Klippert, M. Novello, and J. M. Salim, 
Light propagation in non-linear electrodynamics.
\href{https://doi.org/10.1016/S0370-2693(00)00522-0}{Phys. Lett. B {\bf 482}, 134 (2000)} 
[\href{https://arxiv.org/abs/gr-qc/0005049}{arXiv:gr-qc/0005049}].

\bibitem{obukhov2002}
Y. N. Obukhov and G. F. Rubilar,
Fresnel analysis of wave propagation in nonlinear electrodynamics,
\href{https://doi.org/10.1103/PhysRevD.66.024042}{Phys.\ Rev. {\bf D 66}, 024042 (2002)}.

\bibitem{mignani2017}
R. P. Mignani, V. Tesla, D. Gonz\'alez Caniulef, R. Taverna, R. Turolla, S. Zane, and K. Wu, 
Evidence for vacuum birefringence from the first optical-polarimetry measurement of the isolated neutron star RX J1856.5-3754,
\href{https://doi.org/10.1093/mnras/stw2798}{MNRAS {\bf 465}, 492 (2017)}.

\bibitem{GP99} R. Gambini and J. Pullin, Nonstandard optics from quantum space-time,
\href{https://doi.org/10.1103/PhysRevD.59.124021}{Phys.\ Rev. {\bf D 59}, 124021 (1999)} 
[\href{https://arxiv.org/abs/gr-qc/9809038}{arXiv:gr-qc/9809038}].

\bibitem{ford2013} 
L. H. Ford, V. A. De Lorenci, G. Menezes, and N. F. Svaiter, 
An analog model for quantum lightcone fluctuations in nonlinear optics,
\href{https://doi.org/10.1016/j.aop.2012.10.002}{Ann. Phys. {\bf 329}, 80  (2013)} 
[\href{https://arxiv.org/abs/1202.3099}{arXiv:1202.3099}].

\bibitem{boyd2008} 
R. W. Boyd, 
{\it Nonlinear optics}, 3rd ed. (Academic Press, New York, 2008).

\bibitem{DF19} V. A. De Lorenci and L. H. Ford,  
Subvacuum effects on light propagation, 
\href{https://doi.org/10.1103/PhysRevA.99.023852}{Phys. Rev. A {\bf 99}, 023852 (2019)}
[\href{https://arxiv.org/abs/1804.10132}{arXiv:1804.10132}].

\bibitem{KF18} A. Korolov and L. H. Ford,  Maximal Subvacuum Effects: A Single Mode Example, 
\href{https://doi.org/10.1103/PhysRevD.98.036020}{Phys. Rev. D {\bf 98}, 036020 (2018)} 
[\href{https://arxiv.org/abs/1805.07863}{arXiv:1805.07863}].

\bibitem{bessa2014}
C. H. G. Bessa, V. A. De Lorenci, and L. H. Ford, 
Analog model for light propagation in semiclassical gravity,
\href{https://doi.org/10.1103/PhysRevD.90.024036}{Phys. Rev. D {\bf 90}, 024036 (2014)} 
[\href{https://arxiv.org/abs/1402.6285}{arXiv:1402.6285}].

\bibitem{boyd1972}
G. D. Boyd, E. Buehler, F. G. Storz, and J. H. Wernick,
Linear and Nonlinear Optical Properties of Ternary $A^{II}B^{IV}C_2^{V}$ chalcopyrite semiconductors,
\href{https://doi.org/10.1109/JQE.1972.1076982}{IEEE J. Quant. Elect. {\bf QE-8}, 419 (1972)}.

\bibitem{charra2000} 
F. Charra and G. Gurzadyan, {\it Nonlinear Dielectric Susceptibilities}, in SpringerMaterials, D. F. Nelson, ed. Landolt-B{\"o}rnstein - 
Group III Condensed Matter Volume 30b, Sec. 6.6.1, (Springer-Verlag Berlin Heidelberg 2000).

\bibitem{vahlbruch2008}
H. Vahlbruch, M. Mehmet, S. Chelkowski, B. Hage, A. Franzen, N. Lastzka, S. Go§ler, K. Danzmann, and R. Schnabel, 
Observation of Squeezed Light with 10-dB Quantum-Noise Reduction,
\href{https://doi.org/10.1103/PhysRevLett.100.033602}{Phys. Rev. Lett. {\bf 100}, 033602 (2008)}.

\bibitem{BM69} L. S. Brown, G. J. Maclay, Vacuum Stress between Conducting Plates: An Image Solution, 
\href{https://doi.org/10.1103/PhysRev.184.1272}{Phys. Rev. {\bf 184}, 1272 (1969)}.

\bibitem{SF02} V. Sopova,  and L.H. Ford, The Energy Density in the Casimir Effect, 
\href{https://doi.org/10.1103/PhysRevD.66.045026}{Phys .Rev. D {\bf 66}, 045026 (2002)} 
[\href{https://arxiv.org/abs/quant-ph/0204125}{quant-ph/0204125}]. 

\end{thebibliography}
\end{document}